\documentclass[10pt]{article}

\usepackage{arxivmanuscript}

\hypersetup{
  pdftitle={Surface chemistry investigation of an additively manufactured Al-Mg-Si-Zr alloy: Studies from experiments and first-principles simulation},
  pdfauthor={Zhengqing Wei, Philip Grimm, Inna Plyushchay, Volker Hoffmann, Nebahat Bulut, Lutfi Caglar Ege, Julia Kristin Hufenbach, and Sibylle Gemming}
}

\begin{document}
\justifying

\title{Surface chemistry investigation of an additively manufactured \ce{Al-Mg-Si-Zr} alloy: Studies from experiments and first-principles simulation}

\author[1]{Zhengqing Wei\,\orcidlink{0009-0007-3493-1400}\thanks{Corresponding authors: \href{mailto:zhengqing.wei@physik.tu-chemnitz.de}{zhengqing.wei@physik.tu-chemnitz.de} and \href{mailto:p.grimm@ifw-dresden.de}{p.grimm@ifw-dresden.de}}}
\author[2,3]{Philip Grimm\,\orcidlink{0009-0004-3979-9857}\textsuperscript{*}}
\author[4]{Inna Plyushchay\,\orcidlink{0000-0003-1242-1358}}
\author[3]{Volker Hoffmann\,\orcidlink{0000-0001-8084-6476}}
\author[1]{Nebahat Bulut\,\orcidlink{0000-0002-3087-9154}}
\author[1]{Lutfi Caglar Ege\,\orcidlink{0009-0006-9861-2575}}
\author[2,3]{Julia Kristin Hufenbach\,\orcidlink{0000-0002-1694-2743}}
\author[1,5]{Sibylle Gemming\,\orcidlink{0000-0003-0455-1945}}

\affil[1]{Institute of Physics, Chemnitz University of Technology, Reichenhainer Straße 70, 09126 Chemnitz, Germany}
\affil[2]{Institute of Materials Science, Technische Universität Bergakademie Freiberg, Gustav-Zeuner-Straße 5, 09599 Freiberg, Germany}
\affil[3]{Leibniz Institute for Solid State and Materials Research Dresden, Helmholtzstraße 20, 01069 Dresden, Germany}
\affil[4]{Faculty of Physics, Taras Shevchenko National University of Kyiv, 60 Volodymyrska Street, 01033 Kyiv, Ukraine}
\affil[5]{Research Center MAIN, Chemnitz University of Technology, Rosenbergstraße 6, 09126 Chemnitz, Germany}

\date{}
\maketitle

\begin{abstract}

The surface chemistry of additively manufactured aluminum alloys plays a critical role in corrosion resistance and joining with external materials. In this work, the near-surface elemental composition of a laser powder bed fusion $(\mathrm{PBF\mbox{-}LB/M})$ processed $\mathrm{Al\mbox{-}Mg\mbox{-}Si\mbox{-}Zr}$ alloy was characterized by glow discharge optical emission spectroscopy $(\mathrm{GDOES})$ depth profiling. The measurements reveal pronounced Mg enrichment within the near-surface region extending to approximately $\SI{10}{\micro\meter}$, consistent with the characteristic scale of surface roughness, together with an increase in oxygen concentration, whereas Al, Si, and Zr approach stable bulk-like levels at greater depths. To understand this observation on the atomic scale, first-principles calculations based on density functional theory $(\mathrm{DFT})$ were performed on low-index Al surfaces. The calculated results show a strong thermodynamic driving force for Mg surface segregation, with diffusion energy differences ranging from approximately $\SI{-0.30}{\electronvolt}$ to $\SI{-0.41}{\electronvolt}$, while Zr exhibits a pronounced preference for remaining in the bulk matrix. Vacancy migration calculations further demonstrate that full structural relaxation substantially reduces the migration barrier for Mg to below that of Si, which makes Mg diffusion kinetically highly favorable. Moreover, the presence of adsorbed surface oxygen dramatically promotes the tendency of Mg to diffuse toward the surface, which lowers the diffusion energy difference of Mg to as much as $\SI{-3.0}{\electronvolt}$. This promotes the formation of a locally reconstructed $\mathrm{Mg\mbox{-}O\mbox{-}Al}$ coordinated precursor structure accompanied by localized electron transfer shown by an electron localization function analysis.

\end{abstract}

\vspace{1em}

\section{Introduction}

Additive manufacturing particularly laser powder bed fusion (PBF-LB/M) has been as a powerful technology for producing lightweight aluminum alloy components with complex geometries and tailored mechanical properties \cite{wojciechowski2023additively,hossain2025overview,wang2025data,GEBHARDT2022110796}. Compared with conventional processing routes, PBF-LB/M enables the direct fabrication of intricate three-dimensional architectures that are difficult or impossible to realize using traditional manufacturing approaches \cite{GEBHARDT2022110796,tyagi2023additive}. These complex architectures can be characterized by bicontinuous topology, non-uniform strut thickness, high specific surface area, and tunable anisotropy \cite{kumar2020inverse,vafaeefar2023morphological,alvarez2023mechanical,cai2026nano}.


The surface of such additively manufactured aluminum components represents the primary interface interacting with the environmental surrounding, thereby being crucial to the overall durability and corrosion resistance \cite{linder2024corrosion}. Many material failures originate from surface defects, which makes the nature of the printed surface momentous \cite{linder2024corrosion}. While conventional dense components typically undergo mechanical post-processing to eliminate surface roughness, intricate structures like spinodoid-based metamaterials severely restrict post-processing treatments. However, when considering further surface functionalization, such as the design of metal-polymer hybrid composites \cite{ochoa2011mechanisms} or the application of protective coatings \cite{faure2013dispersion}, the inherent non-flat, rough topography of the as-built state \cite{GEBHARDT2022110796} becomes highly beneficial. It provides enhanced mechanical interlocking and joining sites. Consequently, an understanding of the as-built surface chemistry is important for practical applications.

The surface characteristics of PBF-LB/M parts are expected to differ significantly from those of conventional wrought or cast counterparts, as well as from their raw powders. Although several researchers have characterized the initial powder surfaces \cite{GHASEMI2021102145}, substantial knowledge gaps remain regarding the structural and chemical evolution of the surface immediately after the printing process. PBF-LB/M inherits extremely high cooling rates ($10^{4}$--$10^{6}\ \mathrm{K/s}$) during the cyclic laser thermal fields \cite{bayoumy2023effective,han2023effect}, which inevitably freezes the material into highly non-equilibrium states and promotes the formation of metastable phases at the boundary. This rapid solidification, coupled with aluminum's strong affinity for oxygen, changes the surface and melt-pool boundaries extensively, resulting in a surface layer far more complex than a standard native $\mathrm{Al_2O_3}$ or $\mathrm{AlOOH}$ film, where the latter only occurs when water is present. \cite{Runge2023EnhancingAAO,attia2025enhancing}. Moreover, the elevated surface-to-volume ratio and geometric complexity of spinodoids accelerate oxidation kinetics and induce spatially heterogeneous oxide growth \cite{raza2021degradation,gonzalez2018high}, which can aggravate porosity and micro-cracking at the sub-surface \cite{yin2023nanoscale,nie2024effect}. 

During the repeated heating and cooling cycles that occur in the PBF-LB/M process, solute elements exhibit strong tendencies for surface and near-surface grain boundary segregation, fundamentally driven by atomic size mismatches and chemical reactivity differences \cite{herbig2014atomic,wynblatt2006anisotropy,mahmood2022atomistic}. Among the alloying elements, Mg has a significant solute enrichment at the outermost surface layer during high-temperature exposure \cite{malis1982grain,andolina2021improved,Andreasen2005}. This surface segregation also modulates the initial oxidation kinetics \cite{Panda2009InitialOxidationAlMg}. As oxidation progresses, the initial amorphous alumina film undergoes a complex structural evolution, driven by the reaction of segregated Mg to form $\mathrm{MgO}$ and dominant $\mathrm{MgAl_2O_4}$ spinel phases \cite{Panda2009InitialOxidationAlMg,wu2019oxidation,yoon2020experimental}.

Simultaneously, the diffusion and segregation of solutes including Si and Zr occur at heterophase interfaces or grain boundaries \cite{peng2022influence,lay2016grain}. This changes the diffusion coefficient along these boundaries and thereby affects the mass transport near the surface \cite{pint1998grain}. In Al lattices, Mg and Si diffuse relatively quickly through vacancy mechanisms, whereas transition metals such as Zr exhibit very slow diffusion due to their high solute-vacancy exchange activation energy \cite{simonovic2009impurity}. While Si can influence quenching vacancy concentrations and participate in precipitation sequences \cite{murayama1999pre}, ab initio molecular dynamics (AIMD) simulations indicate a strong chemical affinity between Si and Zr in the melt and at the solidification front \cite{wang2024exploring}. Si atoms are capable of fragmenting large Zr clusters and exerting a drag effect at the solid/liquid interface, which accelerates Zr diffusion and modulates its partition coefficient during solidification \cite{wang2024exploring}. Conversely, the low diffusion coefficient of Zr in Al making it an ideal solute for
resisting recrystallization and refining grain structures  \cite{forbord2004precipitation,elasheri2022improving}. The complex synergy and competition among Mg, Si, and Zr under extreme thermal gradients ultimately affect the multi-component segregation and phase stability of the as-built nanostructures \cite{jiang2021coupled}.

The objective of this study is to investigate the as-built surface chemistry of additively manufactured \ce{Al-Mg-Si-Zr} alloys. Specifically, Glow Discharge Optical Emission Spectroscopy (GDOES) is employed to characterize surface elemental compositions for a sample fabricated by PBF-LB/M. Concurrently, first-principles calculations based on the density functional theory (DFT) are performed to understand the fundamental diffusion tendency toward aluminum surfaces and migration energy barriers toward the surface vacancy for Mg, Si, and Zr alloying atoms in an Al slab, explicitly including the study of the influence of adsorbed oxygen on the surface segregation of Mg.

\section{Methods}

\subsection{Sample fabrication}
A SLM 280 2.0 Dual Laser Machine (Nikon SLM Solutions AG, Germany) was utilized to additively manufacture a cylinder with a diameter of 20~mm and a height of 5~mm. The used powder is made of an \ce{Al-Mg-Si-Zr} alloy (atomizer supplier: NANOVAL GmbH \& Co. KG, Germany) and was further processed by PBF-LB/M with a laser power of 200~W, a scanning velocity of 400~mm/s, a layer thickness of 30~\textmu m, a hatch distance of 100~\textmu m, and a stripe scanning strategy where each layer is rotated by 67°. 

\subsection{Surface analysis}
To investigate the tendencies of various elements to diffuse in the bulk or to remain on the surface, depth profiles were recorded on three different spots by glow discharge optical emission spectroscopy (GDOES) using the device GDA750 HR (Spectruma Analytik GmbH, Germany). A voltage of 800~V and a current of 15~mA were applied in combination with a 2.5~mm universal sample unit. It is worth mentioning, that this material causes flashovers rapidly because Al and Mg generate a high current at constant voltage. To counteract this instability, the current is the regulating variable and kept constant by controlling the voltage. 
For further interpretation, the surface roughness was also determined by measuring the topology via 3D laser scanning microscopy utilizing a Keyence VK-X3000 (Keyence Deutschland GmbH, Germany). 
The extracted values are the arithmetic mean height $R_a$, mean roughness depth $R_z$, and the corresponding area values $S_a$ and $S_z$, according to ISO~25178-2:2021.

\subsection{First-principles computational details}

\begin{figure}[htbp]
\begin{center}
\begin{tikzpicture}
\node[anchor=south west,inner sep=0] (image) at (0,0) {\includegraphics[scale=0.057]{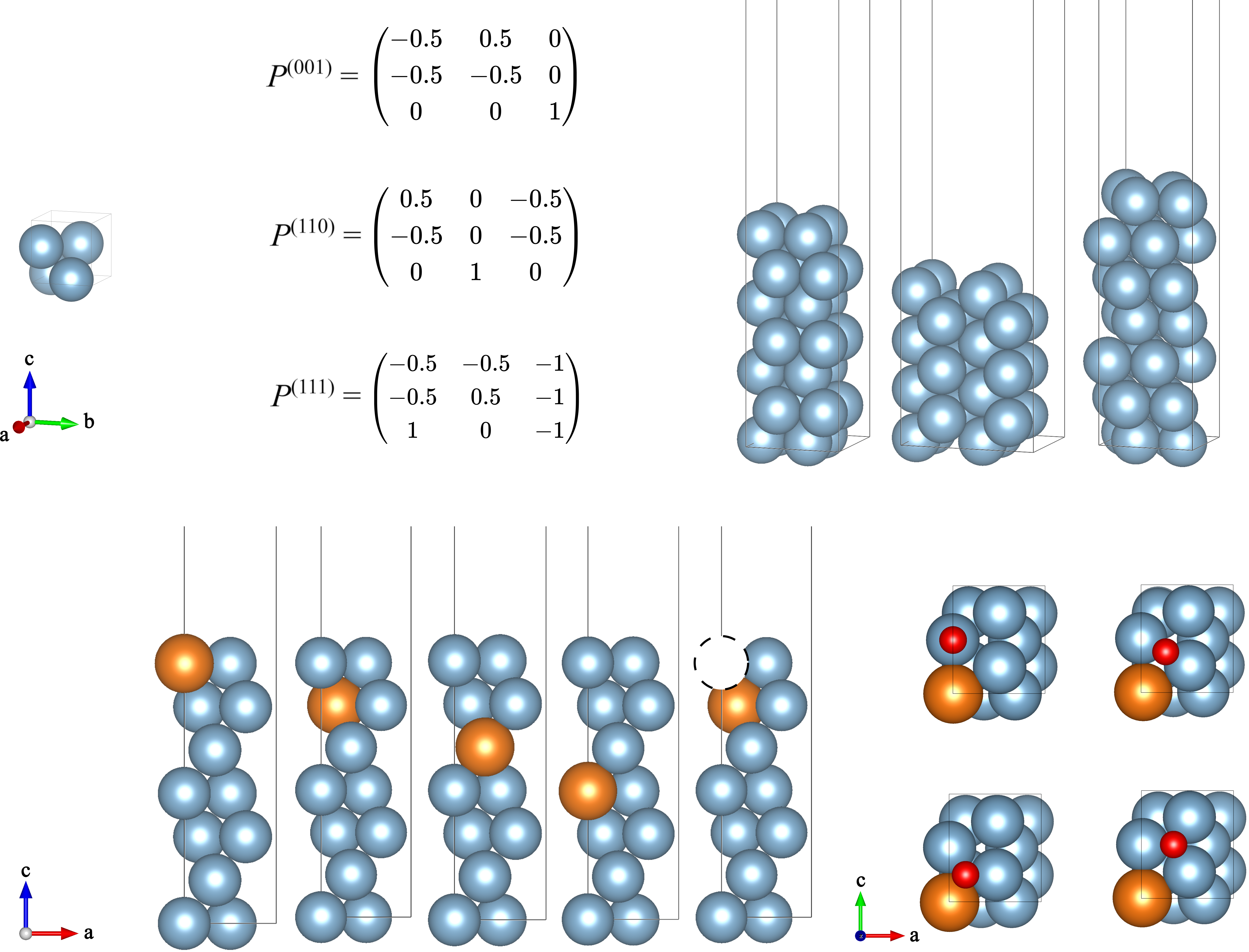}};
\begin{scope}[x={(image.south east)},y={(image.north west)}]
    \node[anchor=south west] at (0.01,1) {(a)};
    \node[anchor=south west] at (0.01,0.4) {(b)};
    \node[anchor=south west] at (0.68,0.4) {(c)};
    \node[black] at (0.05, 0.80) {fcc Al};
    \node[anchor=south west] at (0.23,1) {Transformation Matrix};
    \node[single arrow, draw, fill=white, minimum width=8mm, minimum height= 10mm, anchor=west] at (0.12, 0.75) {};
    \node[single arrow, draw, fill=white, minimum width=8mm, minimum height=10mm, anchor=west] at (0.49, 0.75) {};
    \node[anchor=south] at (0.78,1) {Vacuum thickness of slabs: \SI{30}{\angstromunit}};
    \node[anchor=south] at (0.64, 0.85) {\small (001)};
    \node[anchor=south] at (0.78, 0.85) {\small (110)};
    \node[anchor=south] at (0.92, 0.85) {\small (111)};
    \node[anchor=south] at (0.185, 0.37) {\small Layer 1};
    \node[anchor=south] at (0.293, 0.37) {\small Layer 2};
    \node[anchor=south] at (0.402, 0.37) {\small Layer 3};
    \node[anchor=south] at (0.507, 0.37) {\small Layer 4};
    \node[anchor=south] at (0.615, 0.37) {\small Vacancy};
    \node[anchor=south] at (0.79, 0.4) {\small Top};
    \node[anchor=south] at (0.94, 0.4) {\small Bridge};
    \node[anchor=south] at (0.79, 0.185) {\small hcp};
    \node[anchor=south] at (0.94, 0.185) {\small fcc};
\end{scope}
\end{tikzpicture}
\caption{Structural models of pure Al and surface substituted Al-alloy slabs. (a) Transformation of the face-centered cubic (fcc) Al unit cell into specific crystallographic orientations along the $z$-axis followed by the construction of a $4 \times 7$ surface slab structures.(b) Schematic representation of the Al(111) surface atomic substitution per layer by alloying elements (Mg, Si, or Zr) and the corresponding structure containing surface vacancies.(c) Geometric configurations of the Al(111) alloy slab models under oxygen adsorption with the different high-symmetry adsorption sites.}
\label{fig:1}
\end{center}
\end{figure}
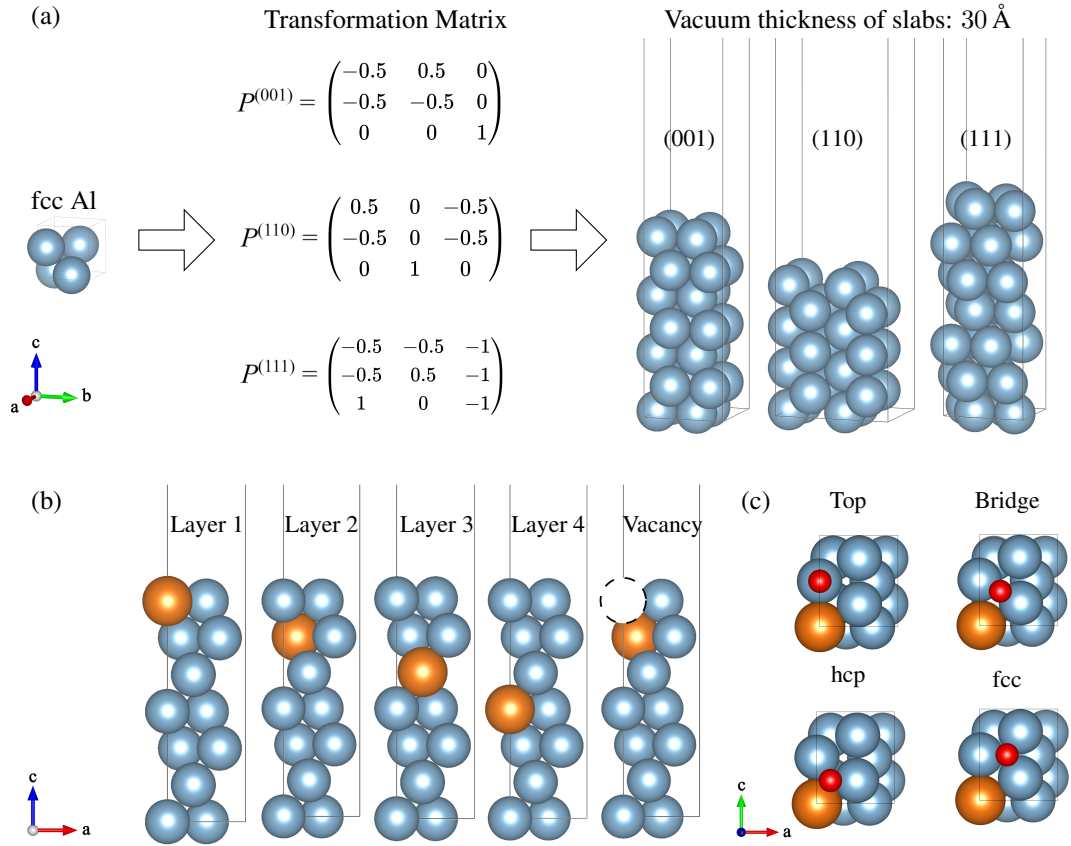

First-principles calculations were performed using the Abinit software package (version 9.10.3) \cite{abinit,abinit1,abinit2,abinit3}, a plane-wave implementation of the density functional theory (DFT), to evaluate the total energies of pure aluminum and its corresponding alloy structures. Electron exchange-correlation effects were treated within the generalized gradient approximation (GGA) using the Perdew-Burke-Ernzerhof (PBE) functional \cite{abinit}. The interactions between valence electrons and ionic cores were described by norm-conserving pseudopotentials. To ensure convergence, the Kohn-Sham electronic wavefunctions were expanded in a plane-wave basis set with a kinetic energy cutoff of 816~eV. Surface structures were modeled using asymmetric slabs consisting of 7 atomic layers (totaling 28 atoms). A 30~\AA\ vacuum layer was introduced along the $z$-direction to eliminate periodic interactions between adjacent slab surfaces. As illustrated in Figure~\ref{fig:1}(a), the (001), (110), and (111) crystallographic orientations were derived from the face-centered cubic (fcc) Al unit cell through appropriate matrix transformations. From these optimized bulk structures, supercells were constructed with dimensions of $2\times2\times3$ for the (001) and (110) planes, and $1\times2\times2$ for the (111) plane.

Geometric optimization was conducted on these slab models by relaxing the nuclear positions of the four uppermost atomic layers, while the bottom three layers remained fixed to simulate the bulk layers. Surface alloy structures, as shown in Figure~\ref{fig:1}(b), were subsequently investigated by substituting surface Al atoms with Mg, Si, or Zr. For all slab calculations, the first Brillouin zone was sampled using a $6\times6\times2$ Monkhorst-Pack $k$-point grid. The electronic self-consistent field (SCF) cycles were considered converged when the total energy difference between iterations fell below $2.72 \times 10^{-6}$~eV. Ionic relaxations were performed until the maximum energy variation between next steps was less than $2.72 \times 10^{-3}$~eV.

\section{Results and Discussion}
\subsection{Experimental results}
\subsubsection{Surface roughness}

\begin{figure}[htbp]
\begin{center}
\begin{tikzpicture}
\node[anchor=south west,inner sep=0] (image) at (0,0) {\includegraphics[scale=0.8]{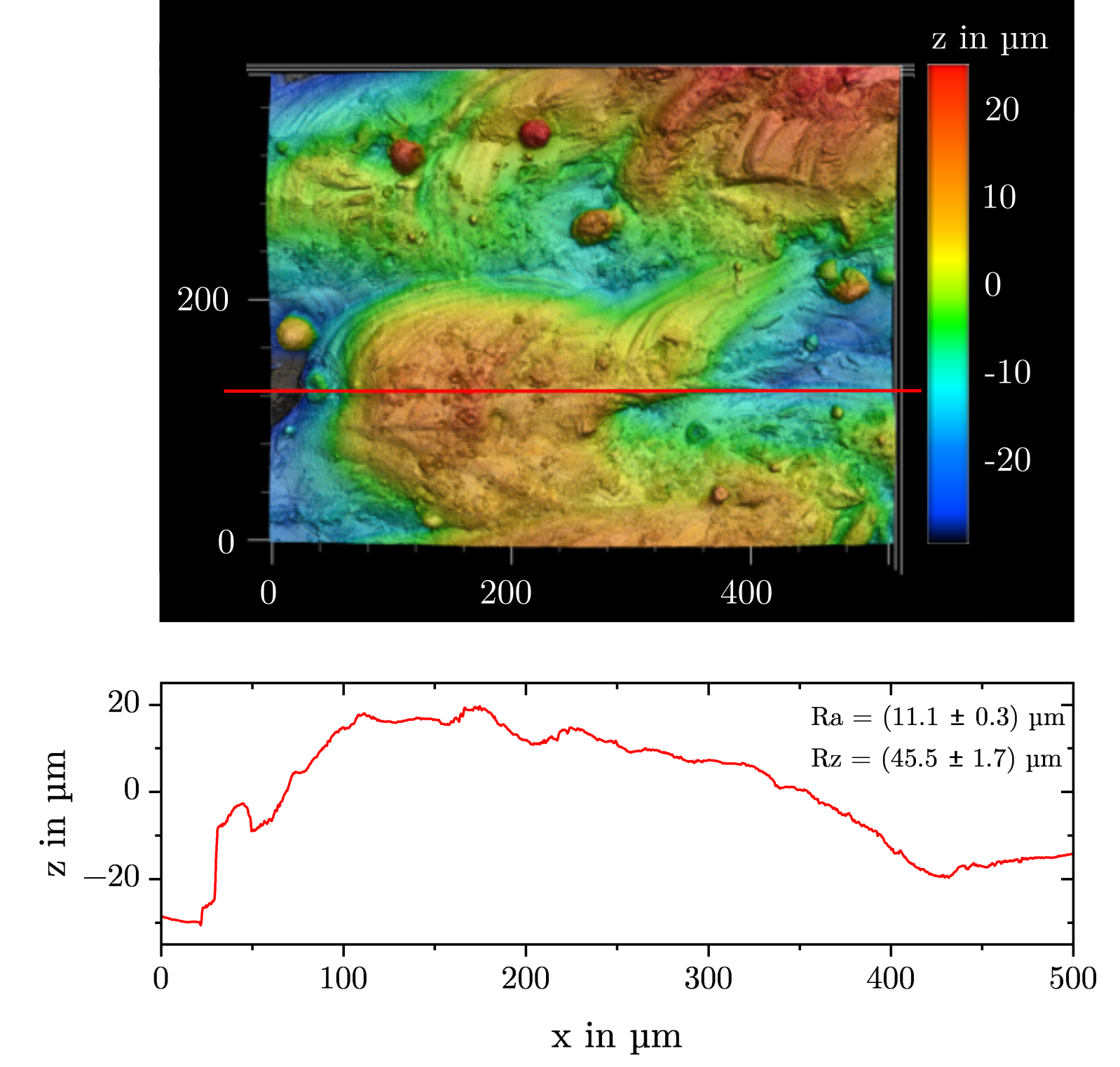}};
\begin{scope}[x={(image.south east)},y={(image.north west)}]
    \node[anchor=south west] at (0.00,0.96) {(a)};
    \node[anchor=south west] at (0.00,0.34) {(b)};
\end{scope}
\end{tikzpicture}
\caption{(a) Topography image by overlaying microscopy image and height profile; and (b) height profile of red line marked in (a) and the determined arithmetic mean roughness $R_a$ and mean roughness depth $R_z$.}
\label{fig:2}
\end{center}
\end{figure}

Additively manufactured parts suffer from poor surface qualities which occur due balling effects, splattering or unstable melt pools during welding \cite{GUNENTHIRAM2018376, GALY2018165}. A typical value of the arithmetic mean area roughness $S_a$ is 9 to 17 \textmu m on the top side of an additively manufactured, commercial AlSi10Mg \cite{ZHOU2021110092}. To discuss the depth profile results achieved by GDOES in the next chapter, it is beneficial to be aware of the surface roughness. In Figure~\ref{fig:2}, the topography and a representative line scan are displayed. The arithmetic mean roughness $R_a$ is (11.1~$\pm$~0.3)~\textmu m and the mean roughness depth $R_z$ is (45.5~$\pm$~1.7)~\textmu m. In addition, the $S_a$ value is 9.4~\textmu m and maximal height $S_z$ is 63.7~\textmu m.


\subsubsection{Elemental depth profile}
\begin{figure}[htbp]
    \centering
    \includegraphics[width=0.75\textwidth]{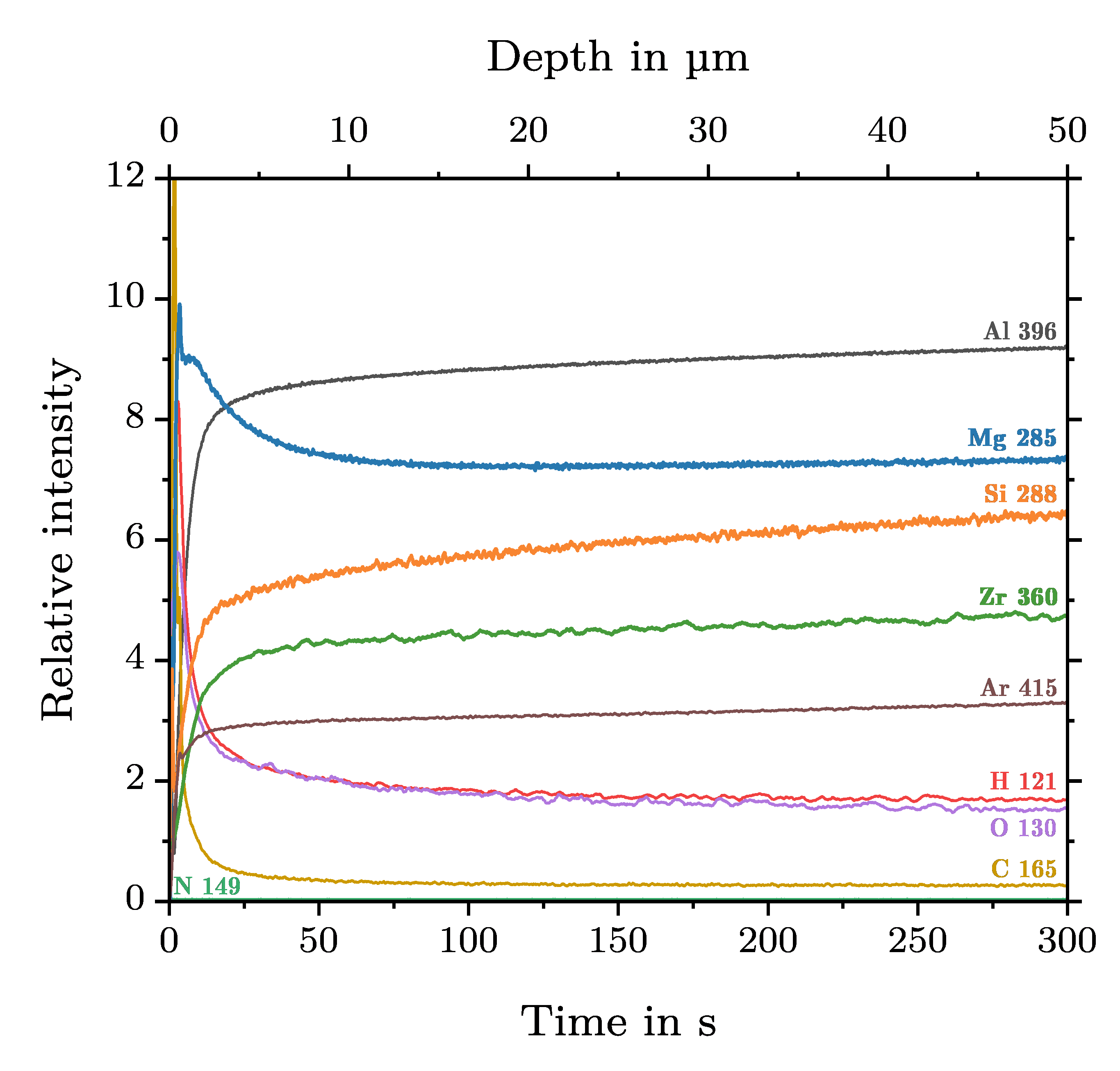}
\caption{Depth profile recorded by Glow Discharge Optical Emission Spectroscopy (GDOES). Curves were smoothed and factorized to facilitate a clearer interpretation.}
\label{fig:3}
\end{figure}
To gain insights about how elements are distributed on the surface and inner core, GDOES was performed. By sputtering continuously the surface and releasing the atoms within the material, the elemental depth profile can be determined, see Figure~\ref{fig:3}. The relative intensity of Al, Mg, Si, Zr, H, C, N, Ar was measured for 300~s which equals a depth of 50~\textmu m. O, C, and H show a maximum in intensity  for the first seconds which dramatically decreases in the first 2~\textmu m. Those elements usually appear due to contamination and moisture on the surface. The following rapid drops are an indicator that the metal surface might already start at roughly 1~\textmu m. Interestingly, the oxygen and hydrogen content drop further permanently and slowly until 40~\textmu m where it reaches an asymptote. Most important, the Mg signal shows also a peak at the beginning and decreases until it reaches the bulk intensity at 10~\textmu m indicating that Mg is concentrated at the surface. Al, Zr, and Si signal is quite low at the beginning and increases until it remains almost constant also at approx. 10~\textmu m, which interestingly is also the size of the surface roughness $R_a$ and $S_a$. Based on these results, it is assumed that Mg forms oxides on top of the surface either as MgO or as a mixed oxide with Al, as spinel $\text{MgAl}_2\text{O}_4$. Generally in the presence of oxygen, Al immediately forms a natural $\text{Al}_2\text{O}_3$ passivation layer of a few nanometers \cite{EVERTSSON2015826}. If exposed for a longer time, the thickness can increase and if anodized it can reach several to hundreds of micrometers \cite{Wernick1987}. However, the thickness of those oxide layers cannot be determined precisely in this work as the surface roughness is around 11~\textmu m, distorting the results which can be observed by slowly stabilizing fade outs.
Ghasemi \textit{et al.}~\cite{GHASEMI2021102145} investigated the powder surface of AlSi10Mg and proposed that the oxide layer composed of Al$_2$O$_3$-MgO-SiO$_2$. Raza \textit{et al.}~\cite{raza2021degradation} reported the formation of predominantly Al$_2$O$_3$ and MgAl$_2$O$_4$ spinel on top of the oxide surface of AlSi10Mg powder after 30 months. The layer thickness was around 4~nm in virgin powder and 38~nm in reused powder. 
However, the enrichment of Mg at the surface after the PBF-LB/M process was not reported in the before-mentioned studies. We assume that the presence of Mg on the surface is a process and/or diffusion related effect. On the one hand, Mg is susceptible to evaporation which also highly influences the weld process~\cite{MALEKSHAHIBEIRANVAND2020106170}. Therefore, the evaporation of Mg within the top layer could be reduced as it is molten and heated up just one time, compared to the underlying layers. However, the Mg peak is located at the first few micrometers which is much smaller than the layer thickness of 30~\textmu m. Nevertheless, a redeposition of Mg in the form of (oxide) particles might be conceivable although a lamellar gas flow and Ar atmosphere is provided. On the other hand, the diffusion of Mg to the surface might be favored due to the high diffusivity and low kinetic barrier. This will be further described in the next chapter by introducing an surface alloy diffusion model.

Furthermore, an additional long sputtering experiment was conducted up to 100~\textmu m to check whether oxygen is incorporated between adjacent melting layers (layer thickness~=~30~\textmu m) during processing. However, an oxygen peak could not be found which means oxygen is not enriched between adjacent layers significantly. This supports the findings from Louvis \textit{et al.}~\cite{LOUVIS2011275} who reported a disruption of oxide interlayers by laser induced vaporization.

\subsection{Calculations}
\subsubsection{Surface alloy diffusion}

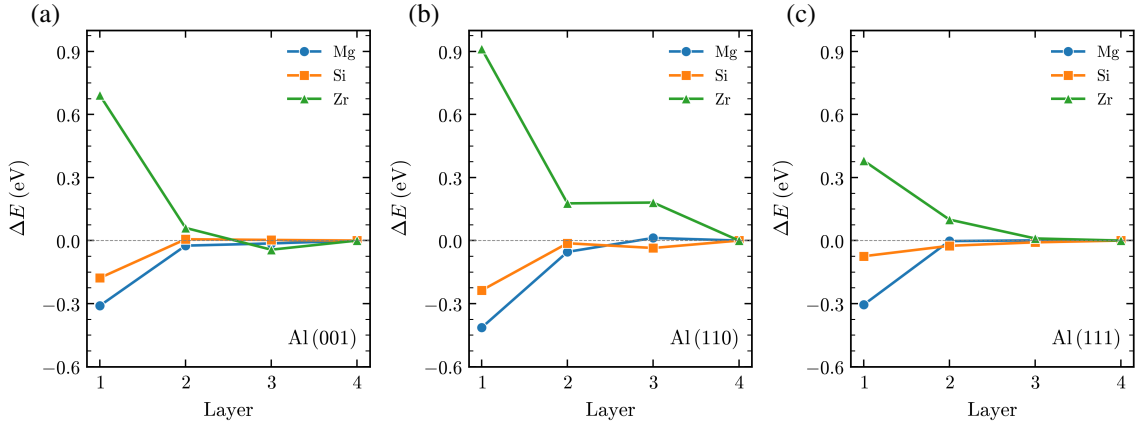
\begin{figure}[htbp]
\begin{center}
\begin{tikzpicture}
\node[anchor=south west,inner sep=0] (image) at (0,0) {\includegraphics[scale=0.54]{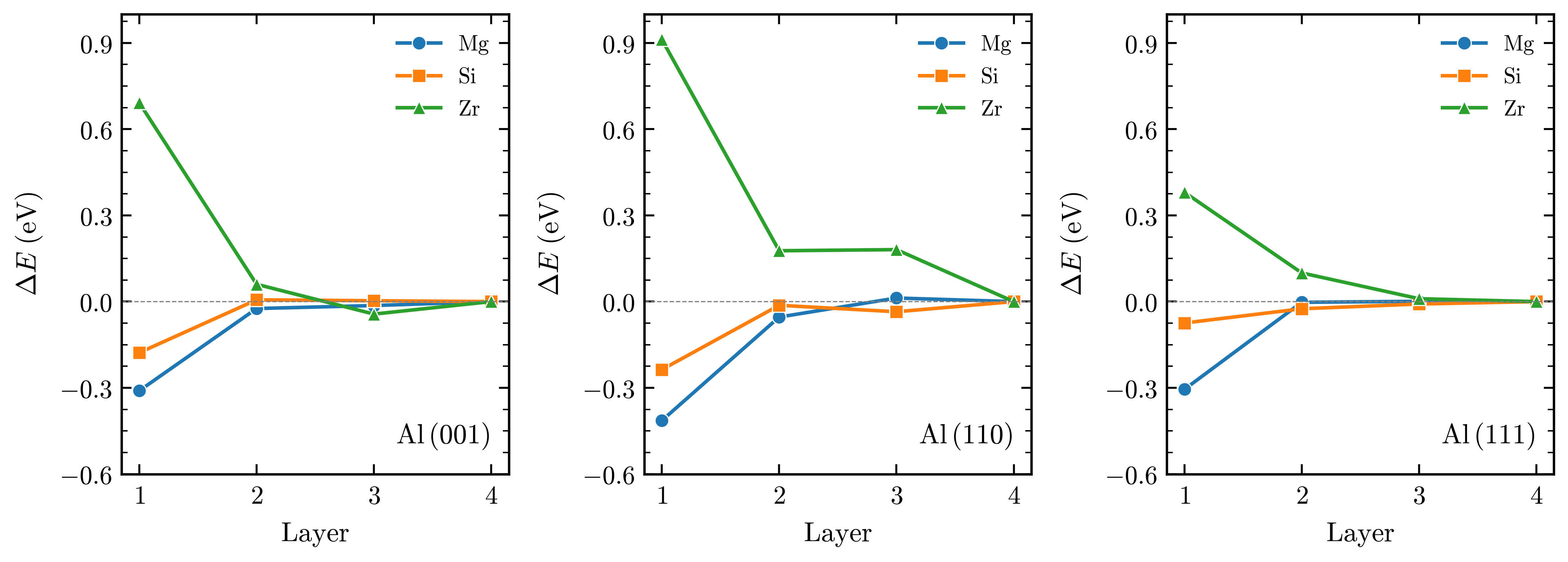}};
\begin{scope}[x={(image.south east)},y={(image.north west)}]
    \node[anchor=south west] at (0.02,0.96) {(a)};
    \node[anchor=south west] at (0.35,0.96) {(b)};
    \node[anchor=south west] at (0.68,0.96) {(c)};
\end{scope}
\end{tikzpicture}
\caption{Energy differences for the diffusion of one alloying atom (Mg, Si, Zr) towards the aluminum surface planes (a) (001), (b) (110), and (c) (111).}
\label{fig:4}
\end{center}
\end{figure}

The experimental results obtained from GDOES raise a fundamental question regarding whether the alloying elements Mg, Si, and Zr preferentially segregate to the surface or remain within the bulk phase. To address this from an atomic perspective, first-principles calculations were performed. Figure~\ref{fig:4} illustrates the diffusion tendencies of these alloying solutes from the bulk matrix toward the aluminum (110), (111), and (001) surfaces. As an alloying atom migrates from a substitutional site within the inner layers to replace an aluminum atom at the surface, the associated energetic variation serves as a thermodynamic indicator of its propensity for surface segregation and its destabilizing effect on the bulk phase. This energy difference of segregation tendency is quantitatively defined as follows:
\begin{equation}
\Delta E = E_{\text{total}}^{\text{upper layer}} - E_{\text{total}}^{\text{deepest diffusion layer}} \label{eq:3.1}
\end{equation}
Here, the deepest diffusion layer corresponds to the fourth atomic layer, which serves as the bulk reference configuration, actually, it is at the central position of the whole slab, which is optimized under the boundary condition of the pure bulk structure by the fixed slab layers below. The first term on the right-hand side denotes the total energy of the slab structure when the solute atom is positioned in a layer above this reference layer. Consequently, a negative value of $\Delta E$ signifies that the alloying element exhibits enhanced stability at that specific upper layer relative to the reference layer. This indicates a thermodynamic driving force for surface segregation. Conversely, a positive value indicates reduced stability at the given layer, which would counteract both in- and out-diffusion. The original way of describing it conveys, that also a positive value leads to out diffusion. Thus, this energetic difference provides a quantitative characterization of the directional diffusion tendencies of alloying elements toward the surfaces.

Figure~\ref{fig:4} compares the trends obtained for the different alloying atoms on the basis of the diffusion energies calculated for the three low-index surfaces. The Mg atom demonstrates a consistent preference across all three surfaces, with negative diffusion energies to the first layer of aluminum ranging from -0.30 eV to -0.41 eV. This indicates that Mg has a strong tendency to reside on the outermost one of these top three planes, as for all three terminations the surface positions are energetically more favorable compared to the interior bulk of aluminum. Additionally, the energy of Mg increases sharply from the first layer to the second layer for all three surfaces, indicating a strong surface segregation trend. Si atoms exhibit slightly negative energies in the surface layer of aluminum ranging from -0.07~eV to -0.23~eV, suggesting a weaker tendency for surface segregation compared to Mg. 

By combining GDOES depth profiling with first-principles calculations, the redistribution of surface atoms during the printing process on Al--Mg--Si--Zr alloys can be interpreted as a multicomponent competitive surface-segregation question that is related to precipitation and oxidation. First-principles calculations indicate that, on idealized and unoxidized Al(001), Al(110), and Al(111) surfaces, Mg and Si possess more favorable driving forces for segregation to the outermost atomic layer than Zr, with Mg exhibiting a stronger segregation tendency than Si. The reference of high-temperature oxidation further shows that even at very low concentrations, Mg preferentially enriches at the alloy/oxide interface during oxidation and undergoes selective oxidation to form Mg-rich oxides such as MgO and MgAl$_2$O$_4$ \cite{karpe2024element}. However, in aged Al--Mg--Si alloys, Si has already been removed from the supersaturated solid solution through the classical precipitation sequence, namely supersaturated solid solution $\rightarrow$ solute clusters/GP zones $\rightarrow \beta''$ (needle-like Mg$_5$Si$_6$) $\rightarrow \beta'/\beta$-Mg$_2$Si, leading to an obvious reduction in the fraction of Si remaining in solid solution \cite{liang2012clustering}. Therefore,  we conclude that Mg- and Si-containing enriched phases are preferentially distributed along grain boundaries or within the bulk. So in Al--Mg--Si--Zr alloys, Mg is more likely to occupy and enrich the outermost surface first, thereby promoting the formation of an Mg-modified oxide layer, whereas Si contributes mainly through the formation of Mg$_2$Si, Si-rich precipitates, and Si-containing inner precipitates or oxides at grain boundaries and within the bulk, rather than through enrichment at the outermost surface. Although first-principles calculations under idealized conditions predict a slightly favorable tendency for Si surface segregation, a pronounced Si-enrichment signal is unlikely to be observed in the near-surface GDOES depth profile of experimental samples. 

In contrast, Zr exhibits markedly positive energy differences in the outermost layer of all three Al surfaces. Specifically, the diffusion energy difference of Zr ranges from $0.38$ to $0.91~\mathrm{eV}$, indicating that Zr atoms tend to avoid the outermost surface because these sites are energetically less favorable than positions in the bulk interior. Moreover, the energy of Zr decreases sharply from the first layer to the second layer on the same surface, further demonstrating that Zr is relatively stable in the subsurface and deeper layers. These calculated results for Zr are in very good agreement with the GDOES measurements, confirming that Zr intrinsically prefers to remain in the bulk region. Studies on high-purity Al reported an activation energy for Al self-diffusion of approximately $1.31~\mathrm{eV}$~\cite{volin1968annealing}. Accordingly, considering that the tracer diffusion activation energy of Zr in Al is about $2.51~\mathrm{eV}$, it is reasonable to conclude that the bulk diffusion of Zr in Al is significantly slower than Al self-diffusion, with a diffusion coefficient approximately three orders of magnitude lower than that of Al self-diffusion~\cite{Marumo1973Diffusion}.

\subsubsection{Surface vacancy migration barrier}

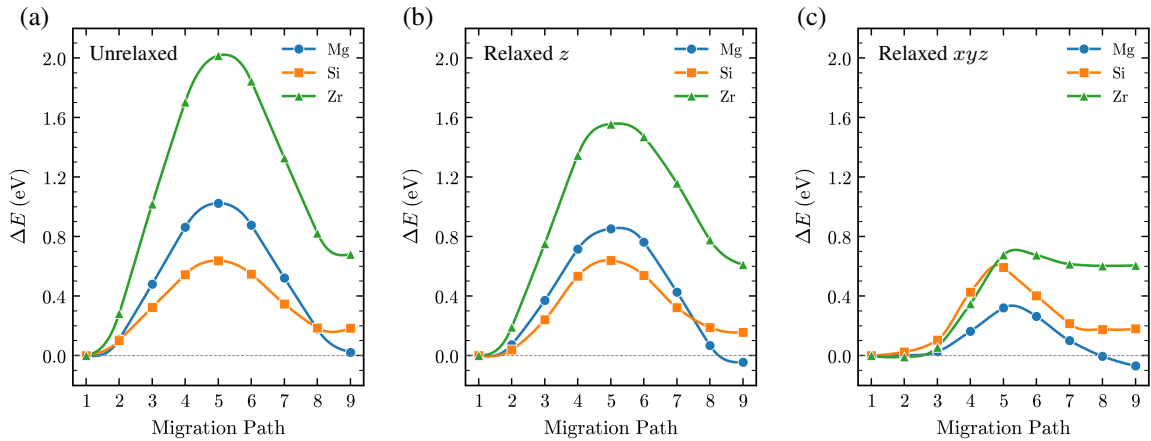
\begin{figure}[htbp]
\begin{center}
\begin{tikzpicture}
\node[anchor=south west,inner sep=0] (image) at (0,0) {\includegraphics[scale=0.535]{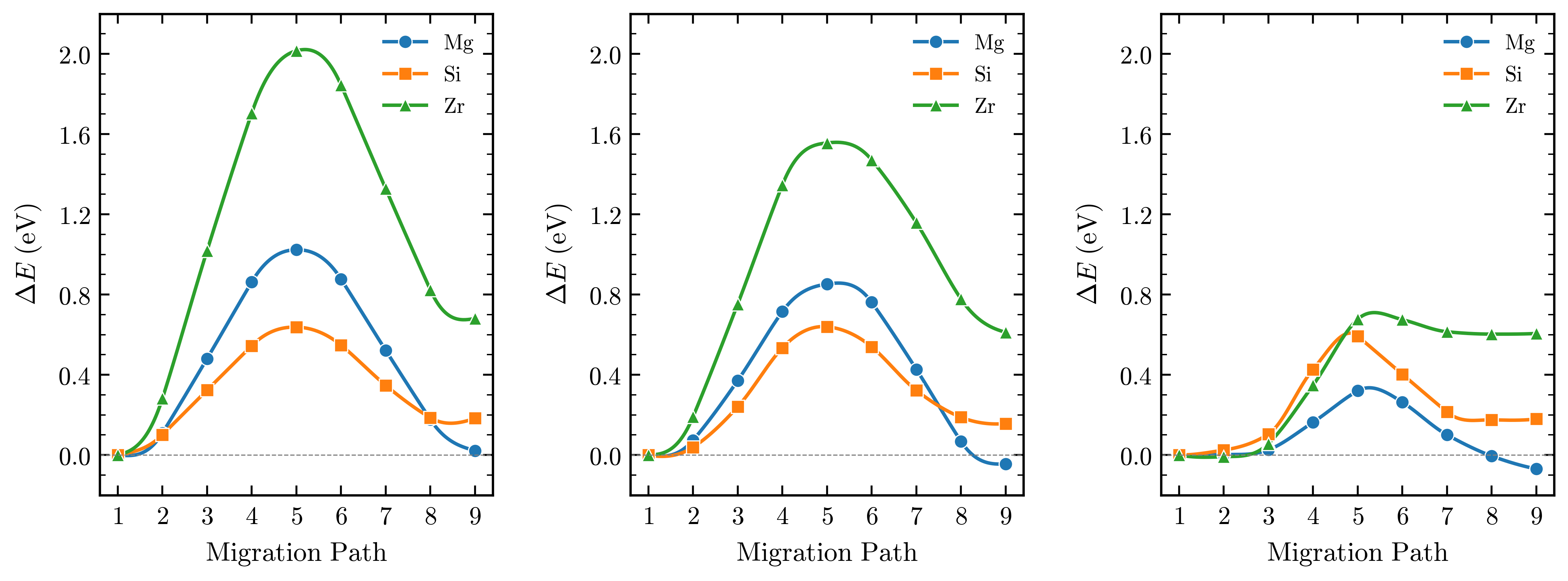}};
\begin{scope}[x={(image.south east)},y={(image.north west)}]
    \node[anchor=south west] at (0.01,0.95) {(a)};
    \node[anchor=south west] at (0.34,0.95) {(b)};
    \node[anchor=south west] at (0.68,0.95) {(c)};
    \node[anchor=south west] at (0.07,0.88) {\footnotesize Unrelaxed};
    \node[anchor=south west] at (0.41,0.88) {\footnotesize Relaxed $z$};
    \node[anchor=south west] at (0.75,0.87) {\footnotesize Relaxed $xyz$};
\end{scope}
\end{tikzpicture}
\caption{Alloying atom migration energy barriers from the second layer to a neighbor vacancy in the first layer on Al(111). The top four surface layers are (a) fixed without relaxation, (b) relaxed only along the $z$-direction, (c) fully relaxed in all cartesian directions. The lines are interpolations and serve as guide for the eye.}
\label{fig:5}
\end{center}
\end{figure}

Although the previous section showed the thermodynamic tendency of Mg and Si to segregate toward the surface, while Zr remains preferentially in the bulk, thermodynamic stability alone is insufficient to determine whether alloying atoms can effectively migrate within the alloy. Surface segregation and diffusion in the matrix are additionally governed by diffusion kinetics, which are controlled by the activation energy barrier associated with vacancy-mediated atomic migration. To assess the kinetic accessibility of surface segregation, the migration barriers for alloying atoms moving from the second subsurface layer to a neighboring surface vacancy on Al(111) were calculated. To understand the influence of local structural relaxation and surface reconstruction on diffusion kinetics, three levels of structural optimizations were considered: the rigid-lattice approximation, relaxation restricted to the surface-normal direction, and full local atomic reconstruction within the top four layers, as illustrated in Figure~\ref{fig:5}. Here, the rigid-lattice approximation reflects the elastic boundary conditions of a pure Al bulk, and the partial relaxation along the surface normal accounts for the elastic background of the pristine Al surface with few defect sites and a low number of segregated atoms. Both reflect the initial stages of the formation of the oxide layer observed in the experiment.

The migration barrier, denoted by $\Delta E$, is defined as
\begin{equation}
\Delta E = E_{\text{total}}^{\text{transition}} - E_{\text{total}}^{\text{initial}}
\label{eq:3.2}
\end{equation}
where $E_{\text{total}}^{\text{transition}}$ is the total energy of the transition state during the migration process, and $E_{\text{total}}^{\text{initial}}$ is the total energy of the initial structure in which the alloying atom occupies the second layer and the vacancy is located at the first layer, as shown in Figure~\ref{fig:1}(b)(Vacancy). A larger migration barrier indicates more kinetically hindered atomic diffusion. In particular, the maximum energy difference at the intermediate position gives an estimate of the migration barrier for each alloying atom.


A comparison of the migration paths in Figure~\ref{fig:5} reveals that the diffusion barriers are strongly dependent on both the alloying species and the degree of structural relaxation. In the unrelaxed configurations shown in Figure~\ref{fig:5}(a), the migration barriers follow the order
$\mathrm{Zr} > \mathrm{Mg} > \mathrm{Si}$. The intrinsic resistance associated with the surface vacancy migration is primarily governed by atomic size mismatch with the Al matrix. Among the three alloying elements, Zr exhibits the largest atomic radius relative to Al\cite{clementi1967atomic}, which introduces severe local lattice distortion during the transition state of the alloy diffusion and vacancy exchange process. Consequently, substantial compressive strain and steric resistance are generated when the Zr atom attempts to migrate from the second subsurface layer into the surface vacancy site, leading to the highest migration barrier of $2.0~\mathrm{eV}$. In contrast, Si possesses a smaller atomic radius\cite{clementi1967atomic} and can more easily pass through the transition configuration with reduced local structural influence, which results in the lowest migration barrier. Mg exhibits an intermediate behavior with its moderate atomic size mismatch.

When relaxation along the surface normal direction is introduced, as illustrated in Figure~\ref{fig:5}(b), the migration barriers of Mg and Zr decrease noticeably. This reduction indicates that vertical structural relaxation partially releases the local strain accumulated around the transition state. Nevertheless, the Zr barrier remains substantially higher than those of Mg and Si, suggesting that relaxation solely along the $z$-direction is insufficient to fully compensate for the strong lattice distortion induced by the Zr atom. Si exhibits only a negligible change of the migration barrier upon $z$-relaxation. We would rather guess that the diffusion process of Si is associated with lower overall strains in the (111) surface layer.

A different behavior appears under full structural relaxation in all three cartesian directions, as shown in Figure~\ref{fig:5}(c). In this case, the migration barriers of Mg and Zr decrease dramatically, whereas the barrier of Si remains nearly unchanged. The strong barrier reduction observed for Mg and Zr demonstrates that lateral atomic displacements and surface reconstruction facilitate the surface diffusion of the alloying atom. In particular, the migration barrier of Mg decreases below that of Si after full relaxation, despite being initially higher in the rigid-lattice approximation. This inversion suggests that Mg diffusion strongly benefits from all cartesian direction relaxation during the migration toward the surface vacancy. The surrounding Al atoms can effectively redistribute the surface strain induced by Mg migration. In contrast, the negligible variation of the Si barrier under all relaxation schemes indicates that the Si diffusion pathway is inherently less related to lattice relaxation effects. As a result, once full surface reconstruction becomes allowed, Mg diffusion becomes kinetically more favorable than Si diffusion.

Another observation shown in Figure~\ref{fig:5} is the energy difference between the initial and final states along the migration pathway. The initial configuration corresponds to an alloying atom located in the second layer with a vacancy at the surface, whereas the final configuration corresponds to the alloying atom occupying the surface site and the vacancy being transferred into the second layer. Thus, the relative energy difference between the two states directly reflects the thermodynamic stability of the alloying atom at the surface in the presence of vacancies. For Zr, the final state consistently possesses a substantially higher energy than the initial state under all relaxation schemes. This indicates that the configuration in which Zr occupies the outermost surface layer is unfavorable. This observation further supports the segregation analysis discussed previously, where Zr was shown to exhibit a strong preference for remaining in the bulk region \cite{wei2025first}. In contrast, the presence of surface vacancies does not suppress the tendency of Mg to diffuse towards the surface. Even after the vacancy-exchange process, Mg remains energetically favorable at the surface region, which is consistent with the Mg enrichment observed in the experimental depth profiles.

\subsubsection{Effect of oxygen adsorption on Mg surface diffusion}

To further investigate the influence of oxygen adsorption on Mg surface segregation, an oxygen atom was adsorbed on the Al(111) surface at four high-symmetry adsorption sites, i.e. the top, bridge, hcp, and fcc positions, as illustrated in Figure~\ref{fig:1}(c). The Mg atom was subsequently substituted layer-by-layer from the fourth layer toward the surface layer. Two structural relaxation ways were considered. In the first way, the lateral elastic boundary conditions of the pristine Al surface at low segregation conditions were maintained and the adsorbed oxygen atom and all atoms within the top four surface layers were allowed to relax only along the $z$ axis, while all lateral coordinates remained fixed. In the second way, full cartesian relaxation was allowed for the nearest-neighbor Al atoms surrounding the substituted Mg atom, thereby allowing partial local surface reconstruction during Mg segregation towards the surface. In both ways, the oxygen atom can relax only along the $z$ axis in order to preserve the selected adsorption configuration.

\begin{figure}[htbp]
\begin{center}
\begin{tikzpicture}
\node[anchor=south west,inner sep=0] (image) at (0,0) {\includegraphics[scale=0.54]{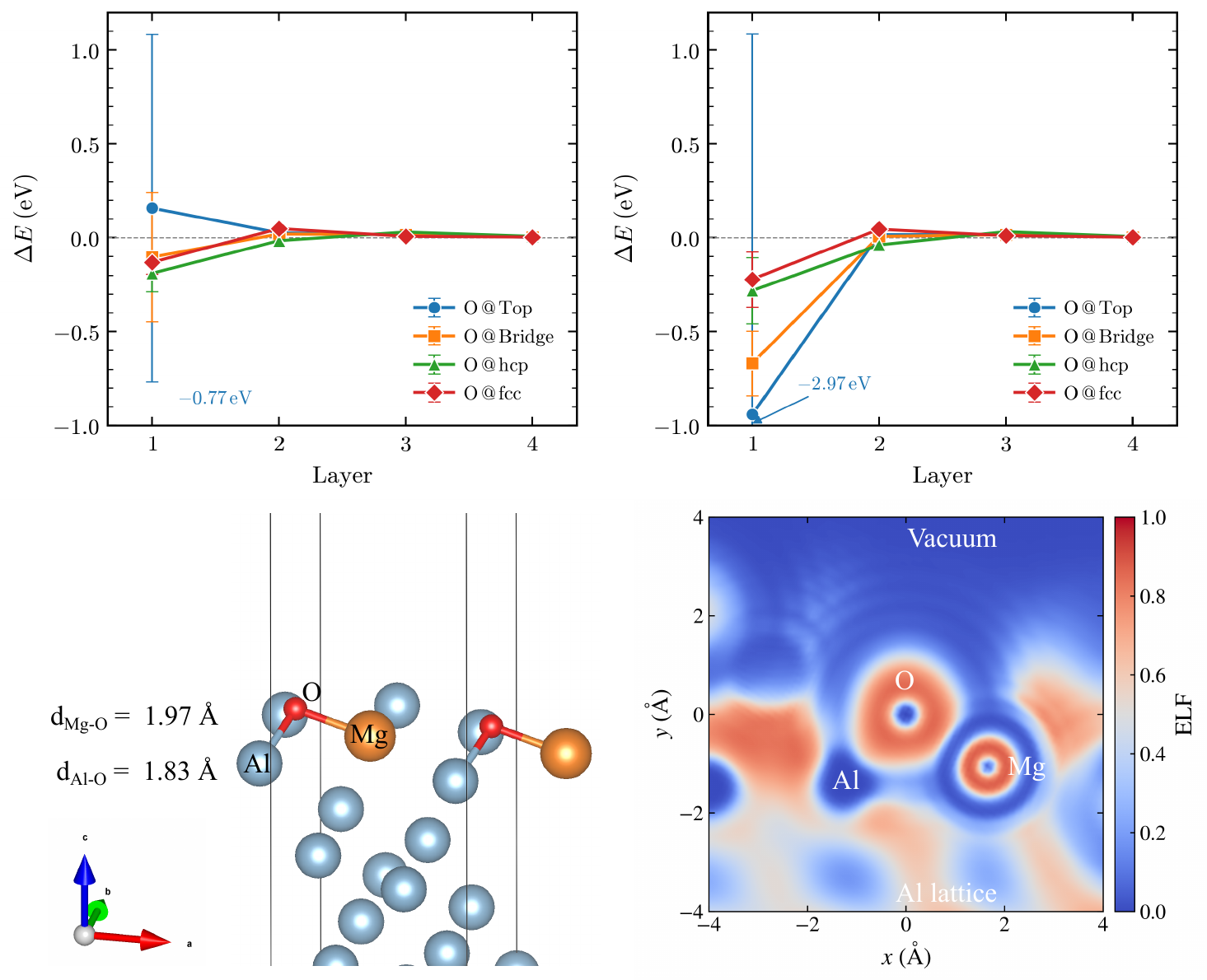}};
\begin{scope}[x={(image.south east)},y={(image.north west)}]
    \node[anchor=south west] at (0.00,0.94) {(a)};
    \node[anchor=south west] at (0.49,0.94) {(b)};
    \node[anchor=south west] at (0.00,0.43) {(c)};
    \node[anchor=south west] at (0.49,0.43) {(d)};
    \node[anchor=south west] at (0.28,0.93) {\footnotesize relax along $z$ axis};
    \node[anchor=south west] at (0.74,0.93) {\footnotesize relax first-shell of Mg};
    \node[anchor=south west, fill=white, font=\footnotesize] at (0.048,0.125) {c};
    \node[anchor=south west, fill=white, font=\footnotesize] at (0.079,0.08) {b};
    \node[anchor=south west, fill=white, font=\footnotesize] at (0.135,0.023) {a};
\end{scope}
\end{tikzpicture}
\caption{Diffusion energy differences of Mg atoms from the fourth layer toward the surface under oxygen adsorption, where the adsorbed oxygen atom and the atoms within the top four surface layers were relaxed (a) only along $z$ axis, (b) using the local first-neighbor relaxation. (c) Optimized atomic configuration corresponding to the energetically most favorable oxygen top-site adsorption structure obtained using the first-shell relaxation model. (d) Electron localization function (ELF) distribution of the optimized structure shown in (c), plotted on the cross-sectional plane passing through the O, Mg, and neighboring Al atom.}
\label{fig:6}
\end{center}
\end{figure}
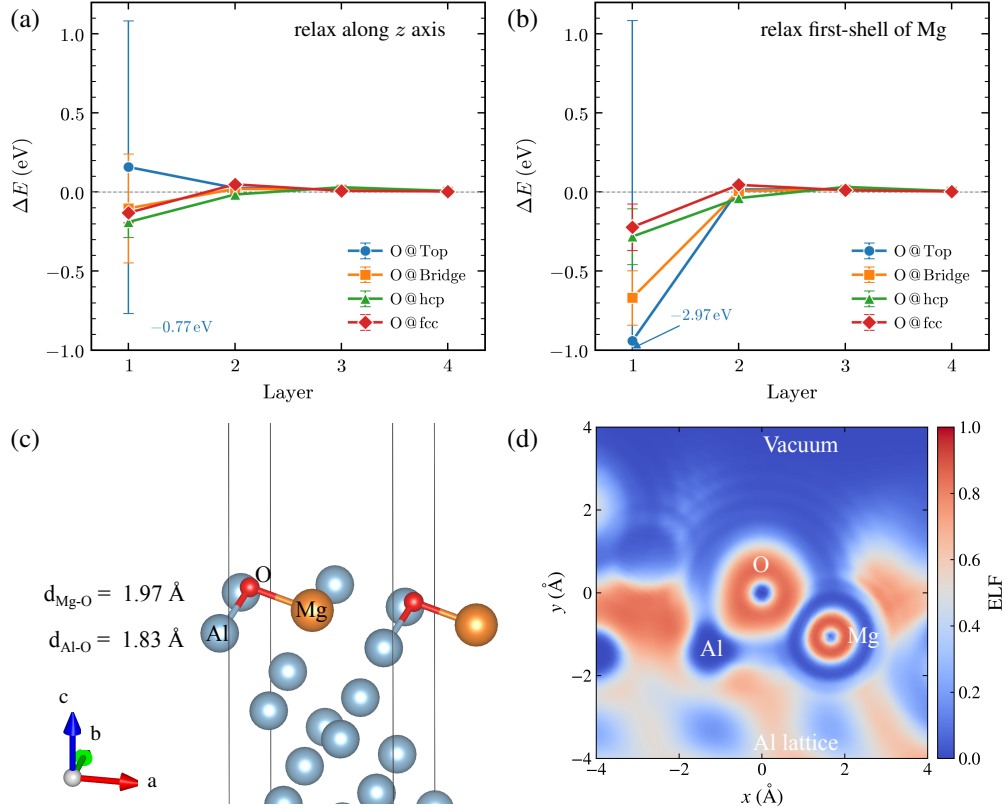

Figure~\ref{fig:6}(a),(b) presents the diffusion energy differences obtained using the two structural relaxation schemes. Overall, distinct energy differences are observed when Mg atoms substitute Al atoms at different surface and subsurface layers. In particular, the largest energy variations occur when Mg occupies the outermost surface layer. For each layer, the plotted data points correspond to the energetically most favorable configuration among all inequivalent Mg substitution sites, while the error bars represent the corresponding energy fluctuation ranges. This configurational dispersion originates from the fact that the adsorbed oxygen atom is constrained to a fixed adsorption site. Consequently, each Mg substitution position generates a distinct local coordination environment relative to the adsorbed oxygen atom.

A comparison between Figure~\ref{fig:6}(a) and (b) reveals significant variations in the segregation tendency of Mg not only among different oxygen adsorption sites, but also between the two structural relaxation schemes. Nevertheless, the energetically most favorable configurations consistently correspond to Mg occupying the outermost surface layer. Compared with the diffusion energy difference of Mg on the oxygen-free Al(111) surface shown in Figure~\ref{fig:4}(c), where the surface segregation energy difference is approximately $-0.3~\mathrm{eV}$, the presence of adsorbed oxygen further stabilizes Mg at the surface, lowering the diffusion energy difference to approximately $-0.8~\mathrm{eV}$ and even down to $-3.0~\mathrm{eV}$ for the oxygen top-site configuration. When partial local reconstruction is allowed as shown in Figure~\ref{fig:6}(b), the stabilization of surface Mg becomes more pronounced. This means lateral relaxation of the top surface atoms plays an important role in accommodating the local Mg--O and Al--O coordinated environment.

Figure~\ref{fig:6}(c) shows the optimized structure of energetically most favorable top-site oxygen adsorbed structure after the first-shell neighboring atoms relaxation. Although the adsorbed oxygen atom remains nearly fixed at its position, pronounced lateral displacements are observed for both the substitutional Mg atom and neighboring surface Al atoms after structural relaxation. This cooperative local reconstruction leads to the formation of an oxygen-centered coordination environment involving one Mg atom and two adjacent Al atoms. The optimized \ce{Mg-O} bond length is approximately 1.97~\AA, while the neighboring \ce{Al-O} bond length is approximately 1.83~\AA. The \ce{Mg-O} bond length is in very good agreement with the values of \SI{1.94}{\angstromunit} reported for normal spinel and of \SI{2.01}{\angstromunit} and \SI{2.04}{\angstromunit} for inverse spinel. Also the \ce{Al-O} bond length is within the range of \SI{1.78}{\angstromunit} to \SI{1.95}{\angstromunit} spanned by normal and inverse spinel structures \cite{mp-3637,mp-38307}. Overall, the values match the inverse spinel structure better than the regular one. The formation of this locally reconstructed \ce{Mg-O}–Al coordinated structure indicates that oxygen adsorption substantially modifies the local bonding environment at the Al surface and energetically stabilizes Mg-containing surface configurations.

To analyze the electronic origin, the electron localization function (ELF) was calculated for this relaxed surface structure as shown in Figure~\ref{fig:6}(d). The ELF distribution reveals strong electronic localization around the oxygen atom, whereas pronounced electron depletion regions are observed around the neighboring Mg and Al atoms. This behavior indicates substantial electronic redistribution induced by oxygen adsorption and suggests a partially ionic bonding character for the \ce{Mg-O} and \ce{Al-O} interactions.

The combined energetic, structural, and electronic analyses consistently demonstrate that oxygen adsorption strongly promotes the stability of Mg-containing surface environments on Al(111). Although the present calculations do not explicitly model oxide growth pathways, the locally coordinated \ce{Mg-O-Al} structures obtained after relaxation may represent precursor configurations associated with the early-stage formation of Mg-enriched oxide layers. Previous experimental studies have reported that selective Mg oxidation on \ce{Al-Mg} alloys can eventually lead to the formation of Mg-rich oxides and MgAl$_2$O$_4$ spinel phases during oxygen exposure \cite{Panda2009InitialOxidationAlMg,wu2019oxidation}. The present first-principles calculations are consistent with this experimentally proposed oxidation tendency and have explained the Mg enrichment observed in the GDOES depth profiles of the additively manufactured \ce{Al-Mg-Si-Zr} alloy.

\section{Conclusion}

This work presents a comprehensive understanding of the surface chemistry and alloying element diffusion of an additively manufactured \ce{Al-Mg-Si-Zr} alloy by combining experimental depth profiling with first-principles calculations. Experimental surface characterization via GDOES depth profiling reveals a distinct near-surface enrichment of Mg within the first \SI{10}{\micro\meter}, which closely coincides with the scale of the arithmetic mean surface roughness ($R_a = \SI{11.1}{\micro\meter}$). In contrast, Al, Si, and Zr concentrations remain significantly low at the outermost surface and stabilize progressively towards the bulk region. Furthermore, the absence of any notable oxygen accumulation between adjacent printed layers indicates that the laser-induced vaporization during the PBF-LB/M process effectively disrupts the continuity of oxide interlayers.

First-principles calculations using DFT provide atomic-scale understanding of the experimental observations from both thermodynamic and kinetic perspectives. The diffusion energy difference analysis confirms that Mg possesses a strong driving force to segregate to the outermost layer of Al(001), Al(110), and Al(111) surfaces, exhibiting negative energy differences which range from $-0.30$ to \SI{-0.41}{\electronvolt}. Conversely, Zr exhibits highly positive diffusion energy differences, which shows its intrinsic preference to remain within the bulk interior rather than migrating outwards. Kinetically, calculations of surface vacancy migration demonstrate that the diffusion pathway is sensitive to surface relaxation schemes and atomic sizes. Upon full three-dimensional structural relaxation, the migration barrier for Mg drops below that of Si, proving that the accumulation of Mg at the surface is both thermodynamically and kinetically favored over other alloying species.

Moreover, the presence of surface-adsorbed oxygen acts as a driver that promotes and stabilizes Mg surface segregation. The chemical interaction with oxygen creates locally reconstructed \ce{Mg-O-Al} configurations with pronounced ionic bonding characteristics, as verified by electron localization function (ELF) analyses. These locally coordinated environments effectively serve as precursor configurations for the initial formation of Mg-rich oxides or spinel phases. Overall, these studies give insights into the surface evolution of \ce{Al-Mg-Si-Zr} alloys during PBF-LB/M processing. In addition, the obtained understanding of surface segregation and oxidation observation may also contribute to the future design and surface functionalization of architected aluminum metamaterials with highly complex geometries, such as spinodoid structures fabricated by PBF-LB/M.

\section*{Acknowledgements}

The authors would like to thank C. Kupka for performing the surface roughness measurements. Furthermore, we gratefully acknowledge the funding provided by the German Research Foundation (DFG) through the Research Training Group GRK 2868: D$^3$ -- Data-driven Design of Resilient Metamaterials (Project No. 493401063).

\section*{Funding}

German Research Foundation (DFG) within the Research Training Group GRK 2868: D$^3$ -- Data-driven Design of Resilient Metamaterials (Project No. 493401063).

\section*{Author contributions}

Zhengqing Wei: Conceptualization, Data curation, Formal analysis, Investigation, Methodology, Software, Validation, Visualization, and Writing -- original draft. Philip Grimm: Conceptualization, Formal analysis, Investigation, Validation, Visualization, Writing -- original draft. Inna Plyushchay: Conceptualization, Data curation, Formal analysis, Investigation, Methodology, Software, Supervision, Validation, and Visualization; Writing -- review \& editing. Volker Hoffmann: Data curation, Investigation, Methodology, Resources, Validation, Writing -- review \& editing. Nebahat Bulut: Methodology and Software. Lutfi Caglar Ege: Methodology and Software. Julia Kristin Hufenbach: Conceptualization, Funding acquisition, Project administration, Resources, Supervision, Writing -- review \& editing. Sibylle Gemming: Conceptualization, Data curation, Formal analysis, Funding acquisition, Investigation, Methodology, Project administration, Resources, Software, Supervision, and Visualization; Writing -- review \& editing.

\section*{Data availability statement}

The data that support the findings of this study are available from the corresponding author upon reasonable request.

\begingroup
\small
\bibliographystyle{unsrt}
\bibliography{util/bibliography}
\endgroup
\end{document}